\documentclass[fleqn,10pt]{wlscirep}
\usepackage[utf8]{inputenc}
\usepackage[T1]{fontenc}
\usepackage{cite}
\usepackage{hyperref}
\usepackage{xcolor}
\usepackage[cal=cm]{mathalpha}
\hypersetup{
    colorlinks,
    linkcolor={red!50!black},
    citecolor={blue!80!black},
    urlcolor={blue!80!black}
}

\newcommand{\au}{\,\mathrm{a.u.}}
\newcommand{\Wcm}{\,\mathrm{W/cm^2}}
 
\newcommand{\HH}{H$_2$\,}
\newcommand{\HHp}{H$_2^+$\,}

\title{Towards simultaneous imaging of ultrafast nuclear and electronic dynamics in small molecules}

\author[1,2]{Saurabh Mhatre}
\author[3]{Zack Dube}
\author[3]{Andr\'e Staudte}
\author[2]{Stefanie Gräfe}
\author[1,*]{Matthias Kübel}
\affil[1]{Institute of Optics and Quantum Electronics, Friedrich Schiller University, 07743 Jena, Germany}
\affil[2]{Institute for Physical Chemistry, Friedrich Schiller University, 07743 Jena, Germany}
\affil[3]{Joint Attosecond Science Laboratory, National Research Council of Canada, Ottawa, Canada}

\affil[*]{matthias.kuebel@uni-jena.de}

\begin{abstract}
When a chemical bond is broken, the molecular structure undergoes a transformation. An ideal experiment should probe the change in the electronic and nuclear structure simultaneously. Here, we present a method for the simultaneous time-resolved imaging of nuclear and electron dynamics by combining Coulomb explosion imaging with strong-field photoelectron momentum imaging. The simplest chemical reaction, \HHp $\rightarrow$ H$^+ +$ H, is probed experimentally for the delay-dependent kinetic energy release, and numerically for the transient change in the photoelectron spectra during the dissociation process. The three-dimensional Schr\"odinger equation is solved in the fixed-nuclei approximation numerically and the results are compared to those from a simple imaging model. The numerical results reflect the evolution in the electron density in the molecular ion as  its bond is first stretched and then brakes apart. Our work shows how simple gas-phase chemical dynamics can be captured in complete molecular movies. 
\end{abstract}
\begin{document}

\flushbottom
\maketitle

\section*{Introduction}

Femtosecond and attosecond laser technology have made it possible to resolve the ultrafast phenomena on the molecular timescales \cite{Zewail2000, Corkum2007, Krausz2009, Lepine2014}. Owing to the substantial difference between timescales on which the electronic and nuclear motion occur, they are usually treated separately in the framework of the Born-Oppenheimer approximation. Although this treatment has proven helpful, it fails in situations where the energy separation of electronic states becomes small, implying slow electronic dynamics. This is the case in the vicinity of avoided crossings and conical intersections \cite{Herzberg1963, Domcke2011}. However, the dynamics of nuclei and electrons are also intertwined even in relatively simple processes, such as bond stretching or bond breakage.

From the experimental side, the correlated motion of electrons and nuclei calls for suitable experimental techniques \cite{Vrakking2014}. Significant progress has been made using, among others, time-resolved fluorescence \cite{Mokhtari1990}, transient absorption \cite{Polli2010}, or high-harmonic \cite{Baker2006, Woerner2010} spectroscopy. However, it is challenging to disentangle the contributions from different molecular processes when detecting only photons. A powerful approach is the use of multi-messenger techniques, such as the coincidence detection of electrons and nuclear fragments \cite{Brehm1967, Davies1999, Doerner2000}. In such experiments, one can, for example, measure the photoelectron spectra associated with different fragmentation channels \cite{Brehm1967, Akagi2009, Boguslavsky2012}, thus identifying the states involved.

Besides the scalar energy values of photoelectrons and ions, spatial resolution is of interest in order to obtain geometrical information about the molecular structure. Imaging of the nuclear geometry of a molecule can be achieved, for example, using Coulomb Explosion Imaging (CEI), where the momentum vectors of multiple ionic fragments from the breakup of a multiply-charged molecule are measured in coincidence \cite{Legare2005, Xu2009}. 
By combining CEI with pump-probe techniques, a series of snapshots visualizing the nuclear motion can be obtained, sometimes referred to as Molecular Movies. Examples include movies of molecular alignment \cite{Rosca-Pruna2001,Karamatskos2019}, vibrational wave packet motion \cite{Ergler2006}, hydrogen migration \cite{Matsuda2011,Liekhus-Schmaltz2015} and roaming dissociation \cite{Ding2019,Tomoyuki2020}. Recently, CEI was combined with coincident photoelectron momentum imaging to follow the ultrafast vibrational wavepacket of \HHp created by tunnelionization from \HH \cite{Hanus2019, Hanus2020}.

For probing the electronic structure, it is promising to turn to the electrons themselves. Using intense laser pulses, the recolliding electron wave packet \cite{Corkum1993} as a means to image molecular structure \cite{Itatani2004,Meckel2008,Amini2020, Amini2021} has received much attention. Laser-induced electron diffraction (LIED) \cite{Zuo1996} allows for measuring the internuclear distances in a molecule if the deBroglie wavelength of the recolliding electron is comparable to the relevant bond lengths \cite{Amini2020, Blaga2012, Wolter2016, Fuest2019, Amini2019}. 
Moreover, the re-scattered wave packet interferes with the unscattered electrons, leading to a holographic interference pattern \cite{Lein2002, Spanner2004, Huismans2011}. This hologram contains information on the rescattering potential \cite{Walt2017}, and the ionization dynamics \cite{Meckel2014, Skruszewicz2015, Haertelt2016, Porat2018}.

In the absence of recollision, the momentum distribution arising from strong-field ionization also contains structural information \cite{Meckel2008, Murray2010, Murray2011, Huang2012, Comtois2013, Li2015}. It was shown in Ref.~\cite{Meckel2008} that the photoelectron momentum distribution (PMD) in the plane perpendicular to the laser polarization resembles the projected valence electron density of the highest occupied molecular orbital. 
Murray et al. showed that orbital imaging by laser-induced tunnel ionization could be understood within a partial Fourier-transform (PFT) method \cite{Murray2010, Murray2011}. Recently, this method was successfully used to reproduce and interpret time-resolved measurements of an electronic wavepacket in the argon cation \cite{Kuebel2019}. However, molecular movies based on laser-induced orbital imaging have not yet been reported.

In the present work, we expand the method put forth in Ref.~\cite{Kuebel2019} to study ultrafast molecular dynamics by combining delay-dependent CEI with orbital imaging through laser-induced tunnel ionization. CEI enables visualization of both bound and dissociating nuclear wave packets, while simultaneously probing the transient electronic structure of the molecular cation in different channels by measuring the PMD in the molecular frame.
Through numerical solutions of the electronic time-dependent Schr\"odinger equation in the frozen nuclei model, we systematically study the dependence of the PMD on the internuclear distance. This translates into a pronounced difference between the PMDs arising from ionization of the bound and the dissociated species, such that they represent an indicator of the rearrangement of the electronic structure during nuclear motion. An intuitive imaging model, which captures the essential details, supplements the rigorous TDSE calculations.

\begin{figure}[h]
    \centering
    \includegraphics[width = 0.5 \textwidth]{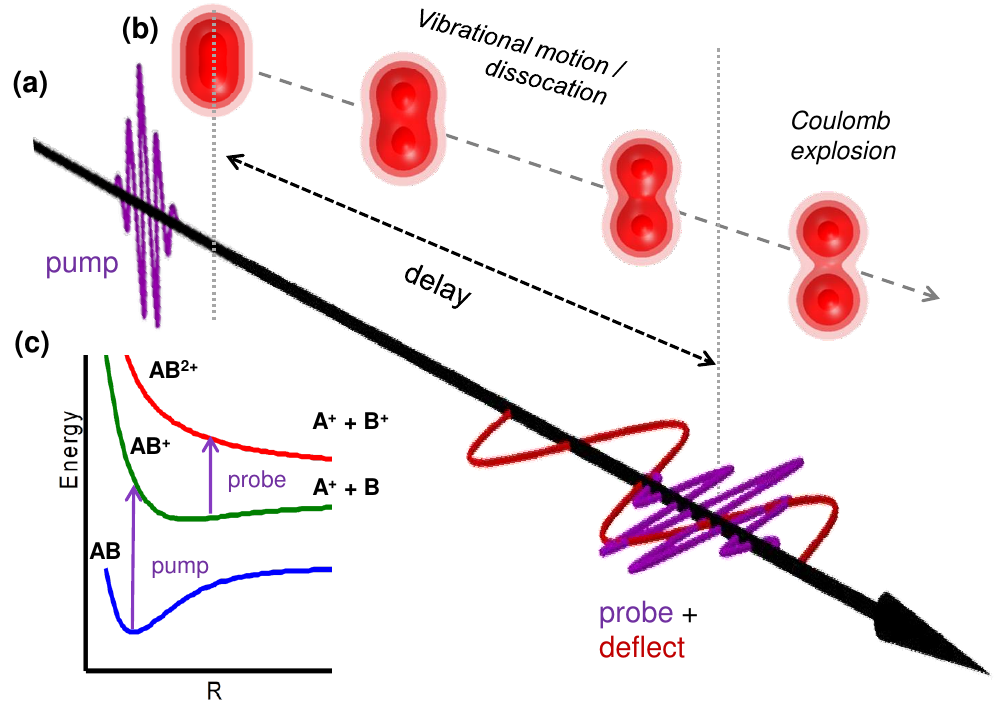}
    \caption{The experimental scheme. (a) A few-cycle pump pulse ionizes the \HH molecule. The induced nuclear and electronic motions in the resulting \HHp ion (shown as a sketch in (b)) are then monitored using the delay-adjustable probe pulse which is perpendicularly polarized to the pump pulse. Along with the probe pulse, a stable mid-infrared pulse is superimposed as a deflection field, which helps in distinguishing the photoelectron spectra generated by the pump pulse and the probe pulse by deflecting electrons generated by the probe pulse. The general ionization pathways in the diatomic molecule are depicted in (c).}
    \label{fig:experimental_scheme_H2}
\end{figure}

\section*{Results and discussion}
\subsection*{Nuclear dynamics}
\begin{figure}[h]
    \centering
    \includegraphics[scale=0.4]{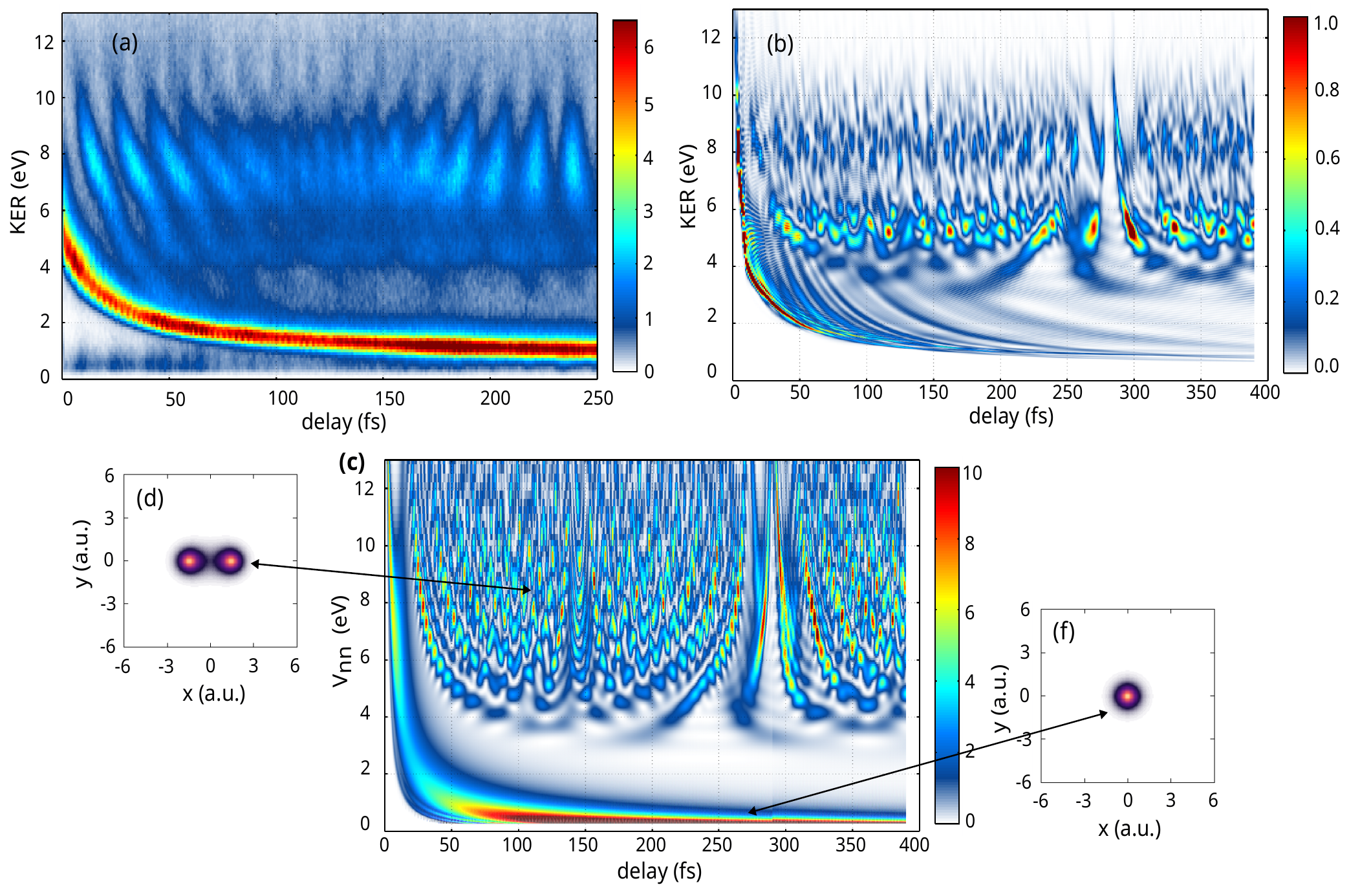}
    \caption{Coulomb Explosion Imaging of the nuclear dynamics: (a) measured KER distribution of protons detected in coincidence, (b) Results obtained with the non-Born Oppenheimer model and considering the interaction with the probe pulse. (c) Calculated evolution of the potential-energy density following interaction with the pump pulse.  The electron density of (d) bound \HHp ion (at $R\approx5\au$), and (e) dissociated molecular ion.}
    \label{fig:CEI_results}
\end{figure}
Experimental and theoretical (non-Born-Oppenheimer TDSE) results for the nuclear dynamics in \HHp, probed by time-resolved CEI, are presented in Fig.~\ref{fig:CEI_results} (a) and (b). To interpret the features observed in the CEI results, the pump-induced time-dependent nuclear density is plotted with respect to the repulsive Coulomb potential of the two protons ($V_{nn}$) in Fig. \ref{fig:CEI_results}(c). The delay-dependent KER distributions reveal two processes taking place after ionization and excitation by the pump pulse: bound vibrational motion, observed at high kinetic energies (KER$\gtrsim 4\,\mathrm{eV}$), and dissociation, observed at low energies.  The KER range covered by the high-energy part in the experimental data, $4\,\mathrm{eV} \leq \mathrm{KER} \leq 10\,\mathrm{eV}$ corresponds to bond lengths in the range of approximately $7.0\,\mathrm{a.u.} \geq R \geq 2.8\,\mathrm{a.u.}$, well above the equilibrium internuclear distance of $R = 2\,\mathrm{a.u.}$, which would correspond to $V_{nn}=13.6\,\mathrm{eV}$. Such large potential energies are clearly present in (Fig.~\ref{fig:CEI_results}(c)) but are not probed at the experimental intensity due to the very large ionization potential at short $R$. The non-Born-Oppenheimer 1D-TDSE results (Fig.~\ref{fig:CEI_results}(b)), show a slightly narrower energy distribution, probably due to a slightly lower intensity used for the calculations.

The delay-dependent KER distributions show pronounced oscillations, which allow us to track the vibrational motion. As shown for D$_2^+$ in Ref.~\cite{Ergler2006}, this type of data allows to accurately determine the composition of the vibrational wave packet contributing to the signal at different KER values. As can be seen in the data, the nuclear wave packet disperses after a few oscillations and becomes delocalized. In the present case of H$_2^+$, after approximately $250\,\mathrm{fs}$, clear oscillations are once again observed as the vibrational wave packet rephases. This behavior is clearly observed in the delay-dependent potential energy distribution of  Fig.~\ref{fig:CEI_results}(c) and also in the non-Born-Oppenheimer 1D-TDSE calculations of Fig.~\ref{fig:CEI_results}(b). The numerical calculations are extended to $400\, \mathrm{fs}$, showing clear revival (rephasing) dynamics at $290\, \mathrm{fs}$.  To investigate these differences, we have performed a Fourier analysis of the experimental and theoretical results. We confirm that the observed dynamics can be attributed to the vibrational transition frequencies of the molecular ion \cite{Ergler2006}. In the case of the full TDSE calculations, additional overtone oscillations are observed. 

The low-energy part (KER$<4\,$eV) is dominated by a narrow distribution, whose central KER value decays with increasing delay. This signal corresponds to the dissociation of the molecular cation. The dissociation channel converges to an energy value of $\approx 1\,\mathrm{eV}$, which is consistent with the kinetic energy gain from the dissociation on the $2p\sigma_u$ state of H$_2^+$ following single-photon excitation from the $1s\sigma_g$ ground state at $R=6\au$. 

The nuclear dynamics, which are probed by the CEI results presented above, are accompanied with variations in the electronic structure. During the bound wave packet motion, the electron density in the molecular ion stretches along the bond. As long as significant density remains between the two nuclei, the chemical bond is intact, as shown in Fig.~\ref{fig:CEI_results}(d). In the dissociated state, the electron density resides on one of the nuclei forming the hydrogen atom, which is depicted in Fig. \ref{fig:CEI_results} (e). Our goal is to access this transition through photoelectron momentum imaging.

\begin{figure}[h]
    \centering
    \includegraphics[scale=0.23]{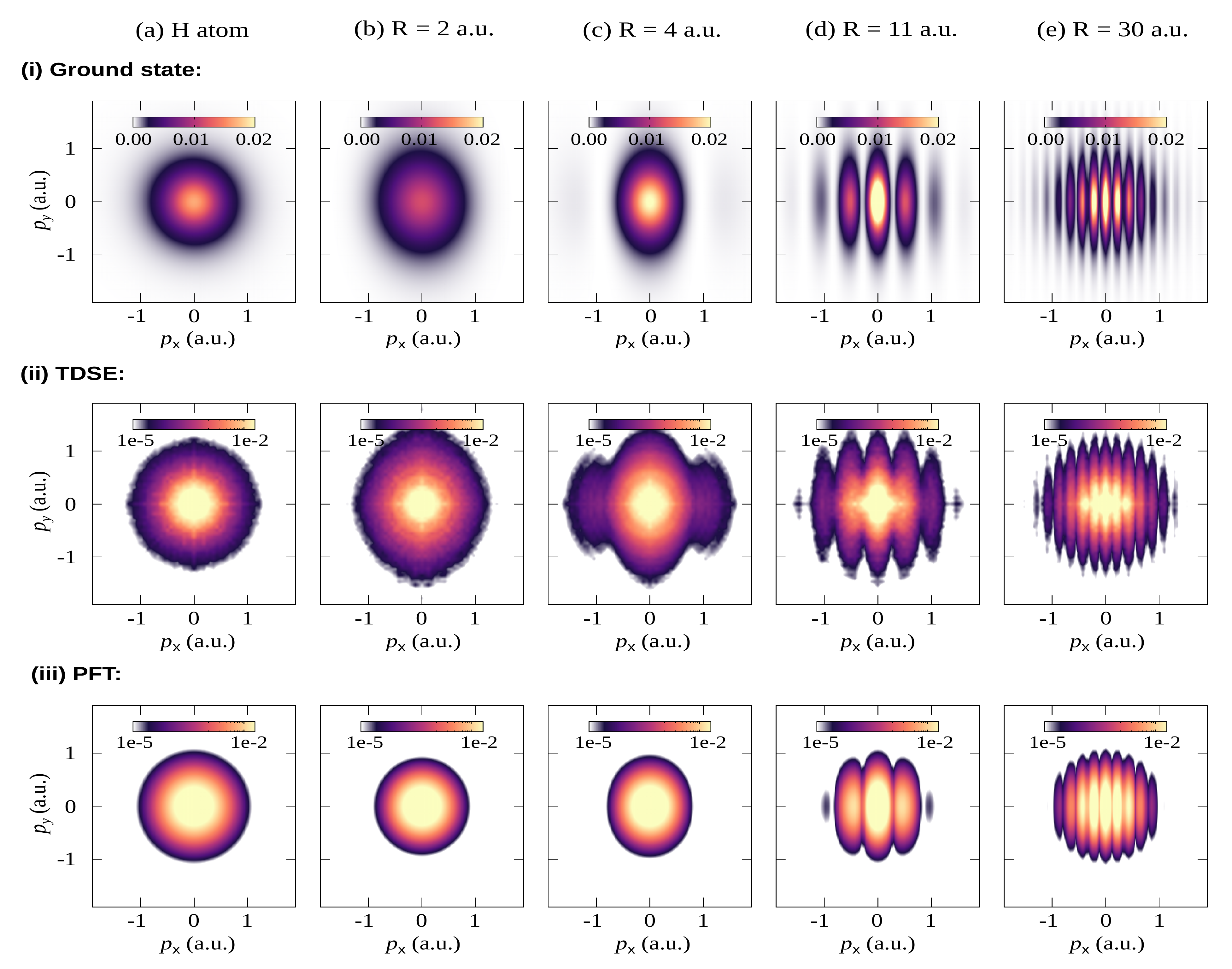}
    \caption{Numerical results for the transversal momentum densities [$\mathcal{M}(p_x,p_y)$], where the (a) hydrogen atom (atomic case) is compared against increasing internuclear distance [(b) - (e)] in the \HHp molecule. The molecular bond is aligned along the $x$-axis. The (i) ground state (initial and bound) electron densities are shown in the upper row, while the photoelectron momentum spectra after the pulse calculated via the (ii) TDSE model and the (iii) PFT model are given in the middle and lower rows, respectively. The color map for the upper row is given in the linear scale, while for the PMDs, the logarithmic scale is used due to relatively low yields.}
    \label{fig:momentum_densities_theory}
\end{figure}
\subsection*{Photoelectron momentum distributions}
The variations of the electronic structure associated with the nuclear dynamics can be probed by direct imaging of the outgoing electron wave packets. In this section, we study the imaging of the electronic density using the 3D electronic TDSE model. Calculations are performed for various internuclear distances and for the hydrogen atom. To evaluate the orbital imprint in the PMD, the three-dimensional momentum space density $|\Phi(p_x, p_y, p_z)|^2$ is integrated over the polarization direction ($z$-axis) yielding the transversal momentum density,
\begin{align}
    \mathcal{M}(p_x,p_y) = \int_{p_z} |\Phi(p_x,p_y,p_z)|^2 \mathrm{d}p_z,
    \label{eqn:momentum_density}
\end{align}
where $\Phi \in [\mathcal{F}\{\Phi_0\}$ (ground state), $\psi^\mathrm{TDSE}_p$ (TDSE calculated PMD), $\Phi^\mathrm{PFT}$ (PFT calculated PMD)$]$. For more details, see the Simulations section. In Fig. \ref{fig:momentum_densities_theory}, the calculated momentum distributions of the bound electron orbital or electronic ground state (upper row) and the PMDs calculated by TDSE (middle row) and PFT models (bottom row) for various internuclear distances are shown. The direct comparison between the momentum distributions for the bound electron and the photoelectron reveals an obvious resemblance between them, albeit the logarithmic color scale used for the PMD due to a weak signal at large momenta. For example, the PMD for the hydrogen atom maintains the circular shape of the orbital. This circular symmetry is lost in the case of the molecular ion, which is narrower along the molecular bond due to the electron density being spread out over the two nuclei. As the bond stretches, a fringe structure emerges, which is also maintained in the PMDs obtained by both TDSE and PFT models. The number of fringes increases with the internuclear distance. The fringe pattern arises due to the interference between electron wave packets emitted from each nucleus, which creates a diffraction pattern similar to Young's double slit experiment \cite{Cohen1966}. 
The diffraction fringes can in fact be used to estimate the internuclear distance using the equation
\begin{align}
    R = \frac{2\pi}{\Delta p_x},
\end{align}
where $\Delta p_x$ is the separation between the two adjacent fringes. These types of double-slit interference in molecular photoionization have been observed in various experiments using intense-laser \cite{Kunitski2019}, XUV \cite{Akoury2007} and x-ray sources \cite{Grundmann2020}.
At large R, the circular shape of the momentum distribution is restored with a fringe pattern superimposed. This indicates that the electron is in a superposition state of being localized on either of the hydrogen nuclei. 

\begin{figure}[h]
    \centering
    \includegraphics[scale=1]{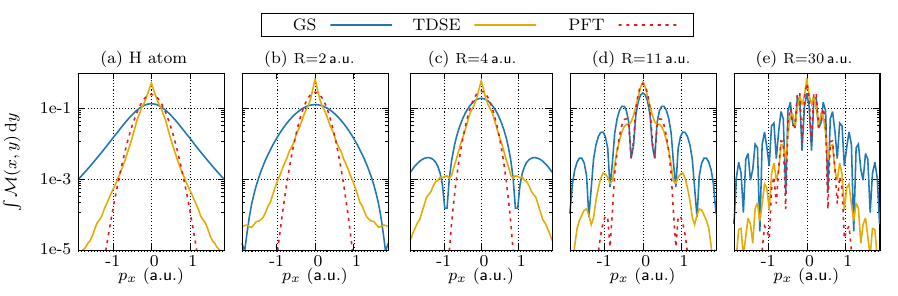}
    \caption{
The momentum distribution along the molecular bond. The momentum distributions of the ground state (GS) wavefunction (solid blue line), and the PMDs obtained using the TDSE (solid yellow line) and the PFT (dashed red line) are calculated by integrating the normalized $\mathcal{M}(p_x,p_y)$ over the $p_y$-axis. }
    \label{fig:momentum_densities_1D}
\end{figure}

For a more quantitative analysis of the momentum distributions, the transversal densities are integrated over the $p_y$-axis, as shown in Fig. \ref{fig:momentum_densities_1D}. This  helps us to focus on the features along the molecular bond. 
In the case of an atom, the momentum distributions diminish monotonously at higher momenta. The decay is superimposed onto the fringe pattern for the molecular case. In comparison to the TDSE results, the PFT results show fewer fringes as the signal diminishing more steeply towards higher momenta. For this reason, the number of fringes appearing in the PFT results (up to a given signal level) is smaller than that for the TDSE results. This discrepancy indicates that the tunnel filter function used in the PFT model overestimates this decay. For the perspective of imaging the electronic density experimentally, this suggests that one may expect more signal at high momentum than predicted by the PFT model. Nevertheless, also for the TDSE results, the photoelectron yield decays much quicker towards higher momenta than the ground state density. 

\subsection*{Analysis with normalized differences}

\begin{figure}[h]
    \centering
    \includegraphics[scale=1]{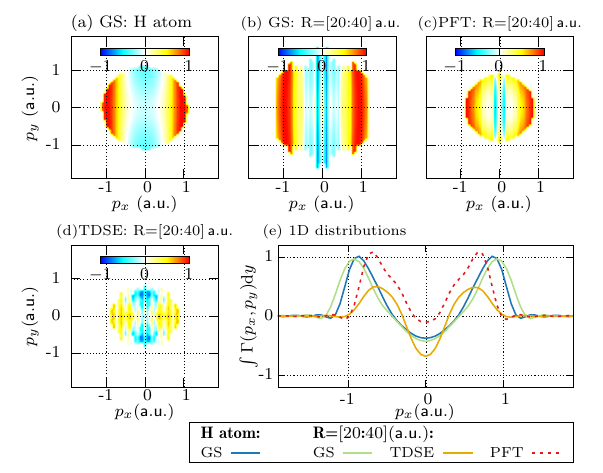}
    \caption{The normalized difference ($\Gamma$) for atomic (a) and atom-equivalent (b)-(d) electron momentum distributions with respect to those of the bound molecule. Here, the corresponding distribution at $R=4\au$ serves as the reference ($\mathcal{M_\mathrm{mol}}$). For the atom-equivalent distributions, the results of the calculations performed for $20\,\mathrm{a.u.}< R < 40\,\mathrm{a.u.}$ were integrated. Results are presented for (a), (b) the ground state (GS) wavefunction, (c) for the PFT model, and (d) the TDSE.  Panel (e) shows a quantitative comparison of the one-dimensional projections of the results presented in panels (a)-(d). The fringes have been removed for clarity by Fourier filtering.} 
    \label{fig:normalized_difference}
\end{figure}

As seen above, the PMDs of the bound and dissociation channels have distinct features along the molecular axis. However, they are dominated by the strong decay of the ionization yield, i.e.~ the effect of the tunnel filter. To suppress the effect of the tunnel filter and reveal the molecular-orbital imprint, normalized difference between two PMDs have been employed in previous work \cite{Meckel2008, Comtois2013,Kuebel2019}. The normalized difference between the transverse momentum distributions ($\mathcal{M}$) of a bound and a dissociated channel is defined as
\begin{align}
    \Gamma(p_x,p_y) = \frac{\mathcal{M}_{\mathrm{atom}}-\mathcal{M}_\mathrm{mol}}{\mathcal{M}_{\mathrm{atom}}+\mathcal{M}_\mathrm{mol}},
\end{align}
where the dissociating (or atomic) PMDs ($\mathcal{M}_\mathrm{atom}$) are compared against the bound (or molecular) PMD ($\mathcal{M}_\mathrm{mol}$).  Figure \ref{fig:normalized_difference} displays the calculated normalized differences. Here, the internuclear distance of $R=4\au$ is considered as the bound channel rather than the equilibrium distance $R=2\au$. This choice is due to the aforementioned low photoelectron yield obtained for the tightly bound molecule. In order to depict the case of the dissociating molecule, we take an average of the momentum densities over the internuclear distance of $R=20\au$ to $R=40\au$. 
The similarity between the normalized difference of the initial momentum densities for the single hydrogen atom, shown in Fig. \ref{fig:normalized_difference}(a), and the dissociated molecule, shown in Fig.~\ref{fig:normalized_difference}(b), confirms the loss of molecular character in the dissociated molecule. 

In Fig.~\ref{fig:normalized_difference}(c) and (d), the normalized differences for the PMDs calculated with PFT and TDSE models are shown, respectively. Compared with the results of Figs.~\ref{fig:momentum_densities_theory} and \ref{fig:momentum_densities_1D}, the normalized differences for the photoelectron resemble the normalized differences for the initial distributions much more closely. This is confirmed by the quantitative comparison in Fig.~\ref{fig:normalized_difference}(e). Thus, normalized differences are in principle suitable to image variations in the electron density. However, as they require the usage of a reference distribution (here: the distribution for the molecular ion at $R = 4 \, \mathrm{a.u.}$) the normalized differences display changes in the electron density rather than absolute densities. The contributions of the two channels (signal and references) can be distinguished in the normalized difference, with the positive values (red) representing dominant contributions from the dissociated atomic orbital, and negative values (blue) representing dominant contributions from the bound molecular orbital. In the present case, the positive values for $|p_x|>0.5\au$, can be attributed to the spherical shape of the atomic orbital, and the reduced widths of the momentum-space orbital of the molecular ion along its bond. Thus, the different widths observed in the normalized differences of  the bound and dissociated channels can be understood as a result of the narrower confinement of the electron when bound to a single atom rather than the molecular ion.


\section*{Conclusion(s) and outlook}

In conclusion, using \HHp as an example, we have investigated a scheme to simultaneously track nuclear dynamics and the associated changes of the electronic structure. The scheme is based on time-resolved Coulomb explosion imaging in combination with photoelectron momentum imaging using intense few-cycle laser pulses. Our approach employs a mid-IR deflection field to separate photoelectron of the neutral precursor molecule from the photoelectron that reports on the molecular dynamics. We demonstrate experimentally that our scheme allows for sufficient time resolution to capture the fast nuclear dynamics of H$_2^+$. The imaging of electronic structure is investigated using computational models, where the electronic 3D TDSE is solved in the fixed-nuclei approximation. The results show a pronounced dependence on the internuclear distance and can thus be used for visualizing the electronic structure changes during the dissociation process. We benchmark the approach of using normalized differences of PMDs to image molecular orbital features and find that the characteristics of bound wave functions are preserved after tunnel ionization. Thus, the proposed scheme offers a practical route to obtain complete molecular movies of both nuclear and electronic dynamics. A refined implementation of the experimental setup is currently in progress.

\section*{Methods}
\label{sec:Methods}
\subsection*{Experimental setup}
The experimental setup is largely identical with the one used previously \cite{Kuebel2017,Kuebel2019}. Briefly, it consists of a femtosecond Ti:Sa laser ($30\,\mathrm{fs}$, $10\,\mathrm{kHz}$, $1.6\,\mathrm{mJ}$) and a Cold target recoil ion momentum spectrometer (COLTRIMS).  The laser output is split into two parts. The first part ($\approx 0.2\,\mathrm{mJ}$) is postcompressed using an argon-filled hollow core fiber, chirped mirrors and glass wedges for dispersion fine tuning. Using a Mach-Zehnder interferometer equipped with a broadband half-wave plate and suitable beam splitters, cross-polarized few-cycle (pulse duration $\tau \approx 5\,$fs full width at half-maximum of the intensity envelope) pump and probe pulses with adjustable time delay are obtained. The dispersion caused by the half-wave plate is compensated in the other interferometer arm using a $1\,\mathrm{mm}$ thick fused silica plate. 
The second part ($\approx 1.4\,\mathrm{mJ}$) of the Ti:Sa output is used to pump an optical parametric amplifier. The resulting CEP-stable mid-IR deflection pulse ($\tau = 40\,$fs, $\lambda = 2300\,\mathrm{nm}$, p-polarization) is recombined with the cross-polarized pump and probe pulses on a Si plate at 70$^\circ$ angle of incidence.  All laser pulses are focused ($f= 75\,\mathrm{mm}$) co-linearly into the center of the COLTRIMS where they intersect a molecular gas jet. The peak intensities of the pump and probe pulses are estimated as $3\times 10^{14}\Wcm$. The intensity of the deflection pulse is $\approx 2\times 10^{13}\Wcm$, and causes no notable ionization on its own. The three-dimensional momentum distributions of ionic and electronic fragments generated in the laser focus are measured using time and position-sensitive detectors.The ion and electron count rates were approximately $5$ and $8\,\mathrm{kHz}$, respectively. This allows for the measurement of clean ion-ion coincidences. However, the electron data is subject to a significant contribution of false coincidences. Further, in the initial implementation of the experiment, which we report here, residual ellipticity in the probe pulse overshadows the orbital imprint in the PMDs, such that the presentation of the experimental results concentrate on the presentation of the data obtained from the ion detector only.  

Fig.~\ref{fig:experimental_scheme_H2}(a) shows the experimental scheme for the interaction with H$_2$. The usage of cross-polarized pump and probe pulses in the two-pulse double ionization scenario helps maximize the number of excited molecules aligned perpendicular to the probe laser polarization. The two generated protons are detected in coincidence with one of the two electrons. Owing to the mid-IR deflection field, those events where the detected electron is produced by the probe pulse, and not by the pump pulse, can be selected with high confidence. The comparison of electron momentum distributions produced by the pump only with those produced by the streaked probe pulse shows that electrons with $p_z < -0.6\au$ are produced almost exclusively by the probe pulse, see also \cite{Kuebel2019}. 

\begin{figure}
    \centering
    \includegraphics[width=0.8\textwidth]{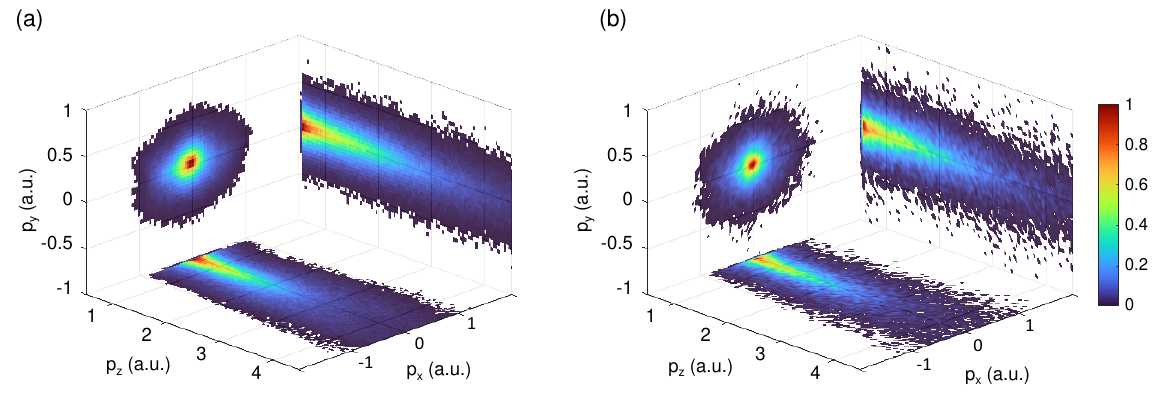}
    \caption{Measured photoelectron momentum distributions for (a) the bound channel with $\mathrm{KER} > 5\,\mathrm{eV}$, and (b) the dissociating channel with  $2.5\,\mathrm{eV} > \mathrm{KER} > 0.5\,\mathrm{eV}$.  }
    \label{fig:ele_exp}
\end{figure}

Three-dimensional momentum distributions of photoelectrons selected for the above condition, and detected in coincidence with ions of different kinetic energies are shown in Fig.~\ref{fig:ele_exp}. The molecular axis is along the x-axis, and the laser polarization is along the z-axis. The fact that the momentum distribution is wider along the x axis than along the y axis is due to a slightly elliptical laser polarization in the experiment.

\subsection*{Simulations}
\subsubsection*{1D nuclear and 1D electron TDSE}
The interaction of molecular hydrogen with the two laser pulses are described within a non-Born Oppenheimer time-dependent Schr\"odinger equation (TDSE) model. Here, the nuclear and electronic motion are co-linear and along the laser polarization (pump and probe pulses). All the following simulations employ atomic units ($\hbar = m_e = e = 4 \pi \epsilon_0 =1$), unless stated otherwise.  The time-dependent Hamiltonian for our model is given as,
\begin{align}
   &H(t) = T_n+ T_e + V_{nn} +V_{ne} + xE(t), &&\\
   &T_n = \frac{p_R^2}{2\mu}, \quad T_e =\frac{p_x^2}{2},&& \\
   &V_{nn} = \frac{1}{R}, \quad V_{ne} = -\frac{1}{\sqrt{(x \pm \frac{R}{2})^2 + \alpha(R)}}
\end{align} 
where $R$ and $p_R$ are the internuclear distance and the corresponding momentum, $x$ and $p_x$ are the electron's distance from the system's center of mass and its corresponding momentum, $\mu$ is the reduced nuclear mass ($918 \, m_e$), $E(t)$ is the time-dependent electric field, and $\alpha(R)$ is the soft-core parameter, respectively. A detailed description of the model can be found in the work of Bandrauk et al. \cite{Bandrauk2009}. 

The dynamics are described in two steps. In the first step, the interaction of the \HHp ion with the pump pulse and the subsequent dynamics is simulated assuming the instantaneous single ionization of \HH near the pulse maximum, i.e.~ a vertical transition from \HH electronic ground state to the \HHp electronic ground state is performed. Thus, the initial state is defined with the Gaussian distribution along the nuclear coordinates centered around $R\approx 1.4 \au$ (\HH equilibrium distance) on the electronic ground state of \HHp ion. The electronic ground state of the \HHp ion is calculated using the imaginary time propagation. In the second step, the interaction with the probe pulse at delay $\tau$ is calculated, where the wave function obtained in the first step after propagation time $\tau$ is used as the initial state. Note that in these simulations, both the pump and probe pulses are polarized along the molecular axis, contrary to the experiment, and the IR deflection pulse is ignored entirely due to the low intensity. The same field parameters are used for the pump and probe pulses, which are of the form, 
\begin{align}
    E(t) = E_0 \sin^2(\pi [t-\tau]/T) \cos(\omega [t-\tau] + \varphi), 
    \label{eqn:electric_field}
\end{align}
with the field strength $E_0=5.14 \times 10^{10} \,\mathrm{V/m}$, angular frequency $\omega=2.511 \,\mathrm{PHz}$ (or wavelength $\lambda=750 \,\mathrm{nm}$), phase $\varphi=0$ and pulse width $T_\mathrm{FWHM}=T/2=5\,\mathrm{fs}$.  These calculations use an equispaced $R$-grid and $x$-grid spanning between $[0:112.5]\au$ and $[-204.8:204.8]\au$, respectively, with 2048 grid points each. The time evolution of the TDSE is carried out using the split-operator method. An absorber (also called mask function) is used to collect the escaped (ionized or dissociated) part of the wave function to avoid reflections off the grid boundaries. The utilized absorber function is described in the ref. \cite{Heather_1987} and reads
\begin{align}
    \zeta_\mathrm{ab}(r;r_a) =  \frac{1}{1+e^{c(r-r_a)}}; \quad c > 0,
    \label{eqn:absorber_function}
\end{align}
where $c$ is the curvature parameter and $r_a$ is the position of the absorber with $r$ being a general grid. Here, for the $x$-grid: $c=0.3$ and $r_a = |150| \au$, and for the $R$-grid: $c=1$ and $r_a = 96 \au$ are used.

The simplest way to follow the pump pulse-induced nuclear dynamics is to assume an instantaneous ionization by the probe pulse, instead of explicitly considering the probe pulse interaction. The kinetic energy release can then be estimated by from the nuclear density on the R-grid, $|\Psi(R,t)|^2 = \int{|\Psi(x,R,t)|^2 \, \mathrm{d}x}$, using the nuclear repulsive potential $V_{nn}$ (see fig \ref{fig:CEI_results}(c)).  However, in order to accurately describe the experimental method, the interaction with the probe pulse followed by the Coulomb explosion is also explicitly simulated. The resulting delay-dependent kinetic energy release (KER) spectra are obtained as follows. The ionized part of the wavefunction is collected using the aforementioned mask function situated near the $x$-grid boundary. This part is then further allowed to evolve in time on the nuclear grid under the nuclear repulsive potential ($V_{nn}$) and the vector field until the mask function on the nuclear grid is reached. The application of mask functions in TDSE for obtaining the photo-induced momentum distribution is described for atoms in ref. \cite{Chelkowski1998} and for molecules in ref. \cite{Bandrauk2009}. 
At each time step the dissociated wavepacket, i.e. the wavepacket reached to the mask function, is Fourier transformed and added coherently to the already dissociated wavefunction $\chi_p(p_R,t)$. Thus, the momentum distribution and corresponding KER spectra of the dissociated nuclear wavepacket at the end of the time propagation are obtained by $|\chi_p(p_R,t_\mathrm{end})|^2$. 

\subsubsection*{3D electronic TDSE with frozen nuclei}
The photoelectron spectra generated by ionization of the \HHp ion by the probe pulse are calculated by solving the three-dimensional electronic time-dependent Schrödinger equation (TDSE) with frozen nuclei. The molecular bond is set along the $x$-axis, while the laser polarization is perpendicular to the molecular bond along the $z$-axis and the field propagates along the $y$-axis. The TDSE for this model under dipole approximation reads
\begin{align}
   &i\frac{\partial}{\partial t} \Psi(x,y,z,t;R) = (T_e + V^\mathrm{3D}_{ne} + zE(t))\Psi(x,y,z,t;R), &&\\ 
   &V^\mathrm{3D}_{ne} = -\frac{1}{\sqrt{(x \pm \frac{R}{2})^2+y^2+z^2}}, 
\label{eqn:TDSE}
\end{align}

where $E$ is the electric field.

Note that the soft-core parameters in the Coulomb attraction potential ($V_{ne}$) are not needed as we consider the complete 3D space and the singularities are taken care of by the finite grid spacing. Thus, these calculations are grid-sensitive. The grid dependency is minimized by ensuring the accuracy of hydrogen atom ground state energies ($I_p = 0.5\au$) produced by the numerical imaginary time propagation calculations on the grid. It is also further validated by reproducing the ionization potential values of \HHp at different internuclear distances reported in the literature \cite{Wind_1965}. The final spatial grid is defined wider and finer along the laser polarization axis ($z$-axis) spanning between $[-102.4:102.4]\au$ with 1024 grid points, while the x and y grids remain coarse with each spanning between $[-51.2:51.2]\au$ with 256 grid points. 

\begin{figure}[h]
    \centering
    \includegraphics[scale=0.9]{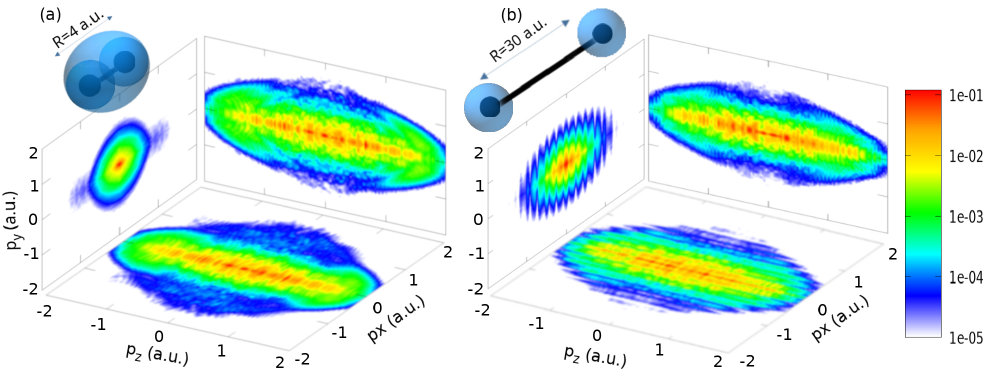}    
\caption{Orthogonal projections of the three-dimensional photoelectron momentum spectra along with the cartoon depictions of respective spatial electron densities for internuclear distances of (a) $R=4\au$ and (b) $R=30\au$.}    
    \label{fig:momentum_density_3D}
\end{figure}

Again, the split operator method is employed for the time evolution of the system, and the ground state $\Phi_0$ obtained by imaginary time propagation is used as the initial state. The escaping wave packets in the ionization direction are smoothly absorbed using an absorber function (eq. \ref{eqn:absorber_function}) near the grid boundaries. In this model, $c=0.6$ and $z_a=\pm 82\au$ are used. The absorber function also acts as a detector to collect the ionized wave packets along the polarization direction (outgoing flux). It is only applied along the laser polarization due to the negligible amount of the wave function reaching the boundaries in the $x$ and $y$ directions. 
The Fourier-transformed outgoing flux is coherently added at each time step and the photoelectron wavefunction at time $t$ is given by,
\begin{align}
    \psi^\mathrm{TDSE}_\mathrm{p} (p_x,p_y,p_z,t;R) = \sum_{t'=t_0}^t \mathcal{F}\left\{ [1-\zeta_\mathrm{ab}(z;z_a)]\Psi(x,y,z,t';R)\right\},
\label{eqn:photoelectron_momentum_wavepacket}
\end{align}
which is further propagated in the vector field without the Coulomb potential. Since the post-ionization propagation is conducted entirely in momentum space,  reflections off the grid boundaries are avoided. Thus, the photoelectron momentum spectra can easily be calculated from the final photoelectron wavefunction densities $|\psi^\mathrm{TDSE}_p|^2$. These 3D momentum densities are shown in Fig. \ref{fig:momentum_density_3D}.  Along the laser polarization ($p_z$) axis, the distribution spreads over a large momentum range, including high energy values ($p_z > 1.5 \au$). However, in this work, our interest lies in the plane perpendicular to the polarization, i.e. the $p_x p_y$ plane. The PMD in this plane retains information on the molecular orbital structure \cite{Meckel2008}. 
The fringe pattern observed in the transverse plane originates from the interference between the nuclei separated spatially along the $x$-axis.

\subsubsection*{Partial Fourier Transform model}
The Partial Fourier Transform (PFT) model \cite{Murray2010,Murray2011} is an intuitive approach to calculate the molecular orbital imprint in PMDs produced by laser-induced tunnel ionization. The central idea is that the transversal density in momentum space at the tunnel exit $z_e$ can be obtained by multiplying the bound wave function at the tunnel entrance $z_0$ with a tunnel filter function. The latter can be derived using the WKB method and making the assumptions of small transversal momenta. It reads \cite{Murray2011}:
\begin{equation}
f(p_x,p_y) = \exp(-0.5 (p_x^2+p_y^2) \tau ),
\end{equation}
where $\tau=\sqrt{I_P / (2 U_P) }/\omega$, with the ionization potential $I_P$, the ponderomotive potential $U_P$, and the laser frequency $\omega$. 

In our implementation, we use the ground state wave functions $\Phi_0(x,y,z)$ obtained by imaginary time propagation as the starting point. A Fourier transform is performed in the plane perpendicular to the laser polarization, and the result is multiplied with the tunnel filter function:
\begin{equation}
|\Psi(z_e,p_x,p_y)|^2 = |\Phi_0(z_0,p_x,p_y)|^2 \cdot f(p_x,p_y),
 \label{tunnel_exit}
\end{equation}
The longitudinal momentum at the tunnel exit $z_e$ is assumed to equal zero.
To obtain a three-dimensional momentum distribution, the acceleration in the laser field is described as a convolution with a Gaussian function,
\begin{equation}
|\Phi^\mathrm{PFT} (p_x,p_y,p_z)|^2 =\frac{1}{2\pi w_z} |\Phi(p_x,p_y,0)|^2 \ast \exp{\left(-\frac{p_z^2}{w_z^2}\right)}, 	
 \label{eq:streaking}
 \end{equation}
where $w_z = 1\au$. 


\bibliography{main}

\end{document}